\documentclass{aa}
\usepackage[varg]{txfonts}
\usepackage[utf8]{inputenc}
\usepackage{comment}
\usepackage{amsmath}
\usepackage{graphicx}
\usepackage{gensymb}
\usepackage[colorlinks=true,linkcolor=blue,citecolor=blue]{hyperref}
\usepackage{xspace}
\usepackage{geometry}  % Adjust the page layout if needed
\usepackage{lipsum}    % For generating sample text (optional)

\usepackage{gensymb}

\newcommand{\sirocco}{\textsc{Sirocco}\xspace}

\newcommand{\kbolx}{$k_\mathrm{bol,X}$\xspace}

\usepackage{orcidlink}

\usepackage{natbib}
\bibpunct{(}{)}{;}{a}{}{,} % to follow the A&A style

\begin{document}

\title{Quenching of X-ray emission in little red dots \\by both Compton-thick gas and high accretion rates}
\titlerunning{Quenching of the X-ray emission in LRDs by Compton-thick gas and high accretion rates}

\authorrunning{Sneppen et al.}

\author{Albert Sneppen\inst{\ref{addr:DAWN},\ref{addr:jagtvej}}\orcidlink{0000-0002-5460-6126}, 
Darach Watson\inst{\ref{addr:DAWN},\ref{addr:jagtvej}}\orcidlink{0000-0002-4465-8264},
James H. Matthews\inst{\ref{addr:oxford}}\orcidlink{0000-0002-3493-7737}, 
Stuart A. Sim\inst{\ref{addr:belfast}}\orcidlink{0000-0002-9774-1192} \\
}

\institute{Cosmic Dawn Center (DAWN)\label{addr:DAWN}
\and
Niels Bohr Institute, University of Copenhagen, Jagtvej 128, DK-2200, Copenhagen N, Denmark \label{addr:jagtvej} 
\and
Astrophysics, Department of Physics, University of Oxford, Keble Road, 
Oxford OX1 3RH, UK\label{addr:oxford}
\and
School of Mathematics and Physics, Astrophysics Research Centre, Queen's University Belfast, Belfast, United Kingdom\label{addr:belfast}
}

\date{Received date /
Accepted date }

\abstract{ Little red dots (LRDs) are candidate high-redshift supermassive black holes accreting in dense gas. They remain undetected in X-rays. In previous work, we provided the first quantitative models that reproduce the optical and near-infrared spectra of LRDs with the \sirocco radiative transfer code, thereby constraining the properties of the surrounding gas. Here, we use these constraints to predict the X-ray attenuation produced by dense gas cocoons, and explore its dependence on Balmer-break strength, metallicity, intrinsic X-ray spectral energy distribution, and observed bandpass as a function of redshift. The X-ray constraints are very tight, requiring extinction by a Compton-thick gas column (\(N_{\rm H}\sim10^{25}\)\,cm\,\(^{-2}\)) with moderate metallicity (0.05--0.1\,\(Z_\odot\)) and intrinsically weak X-ray emission (the ratio of bolometric to X-ray luminosity is \(k_{\rm bol,X}\gtrsim 30\)), as observed in narrow-line active galactic nuclei with high accretion rates, to make LRDs sufficiently faint to evade detection. Intrinsically bright X-ray emitters as seen in typical broad-line active galactic nuclei would be detected even behind the typical Compton-thick gas columns with modest metallicity that were inferred from the optical spectra. Very low metallicity objects might be detected in X-rays even with low intrinsic X-ray luminosities, suggesting that LRDs are not (currently) chemically pristine.
    }

\maketitle

\section{Introduction}

X-rays are a powerful diagnostic of active galactic nucleus (AGN) activity \citep{Ueda2014}. However, the putative AGN nature of little red dots (LRDs) identified by JWST at high redshifts is challenged by their non-detection in deep \emph{Chandra} X-ray data. The bolometric luminosities of LRDs, $L_{\rm bol}\!\sim\!10^{43}{-}10^{45}\,\mathrm{erg\,s}^{-1}$, were inferred from the rest-frame optical emission \citep{Greene2025,deGraaff2025b}. For typical AGN bolometric corrections, this is in tension with their non-detection in deep X-ray measurements, with upper limits of $L_{\rm 2{-}10\,keV}\lesssim(2{-}4)\times10^{41}\,\mathrm{erg\,s}^{-1}$ \citep{Maiolino_Chandra_2024,Ananna2024,Yue2024,Sacchi2025,Comastri2026}.

The high surface brightness and compact sizes of LRDs suggest AGN accretion \citep{Fujimoto2022,Kokorev2024,deGraaff2025,Yanagisawa2026}, as do their broad emission lines \citep{Matthee2024,Greene2024,Juodzbalis2025}. However, LRDs probe a novel radiative-transfer regime: dense gas, in which a Balmer break shapes the optical continuum \citep{Inayoshi2025,Naidu2025,deGraaff2025}, and electron scattering, rather than bulk kinematics, dominates the line widths \citep{Rusakov2025,Chang2025,Nikopoulos2025}. After correcting for electron scattering, the characteristic black hole (BH) mass is $M_{\rm BH}\sim10^{6} M_\odot$, which implies accretion at close to the Eddington limit \citep{Rusakov2025,Sneppen2026}.
In particular, \cite{Sneppen2026} showed that AGN emission mediated through optically thick non-spherically symmetric dense gas flows around a supermassive BH would unify the diverse population and simultaneously reproduce the ensemble of optical and near-infrared observables. The photoelectric absorption and scattering from such Compton-thick gas columns is expected to depress X-ray emission \citep{Juodzbalis2024_rosetta,Maiolino_Chandra_2024,Rusakov2025}. The extent to which the LRD X-ray weakness is driven by extreme accretion conditions (e.g.\ suppressed coronal emission) or by heavy obscuration, however, remains uncertain \citep{Tortosa2026}. In this work, we build on the models of \citet{Sneppen2026b,Sneppen2026} to quantify X-ray attenuation in the dense gas cocoons likely surrounding the early-phase supermassive BHs seen in LRDs and to test hard and X-ray-bright coronae versus soft and X-ray-weak coronae.

\begin{figure}
\vspace{-0.1cm}
\begin{center}
    \includegraphics[angle=0,width=0.5\textwidth]{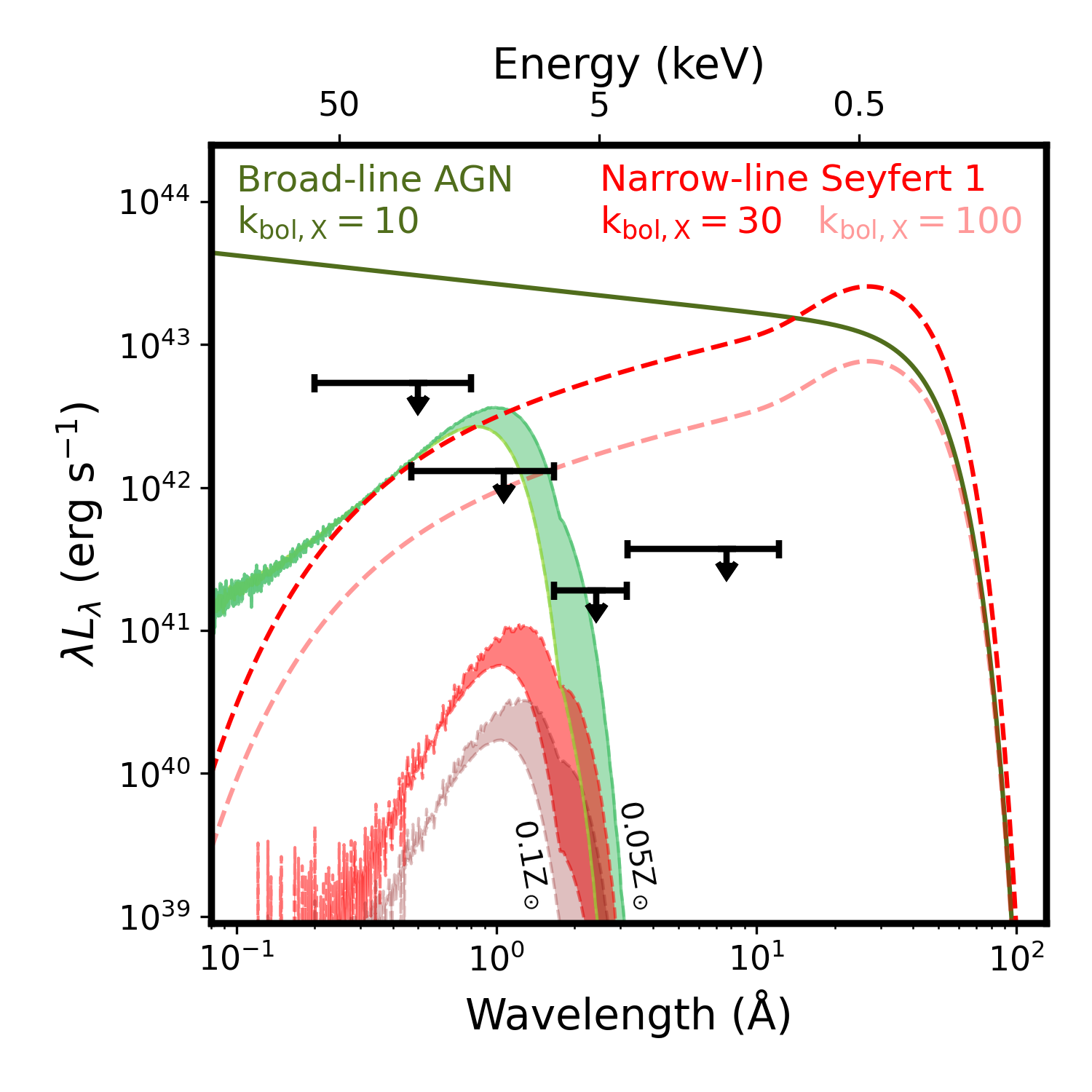}
\end{center}
\vspace{-0.9cm}
\caption{X-ray SEDs of LRDs modelled with \sirocco. The X-ray upper limits for JWST AGNs in the \emph{Chandra} Deep Field North and South are shown in black \citep{Sacchi2025,Comastri2026}; see Fig.~\ref{fig:type2} for type~2 AGN constraints. The intrinsic X-ray templates are shown for (i) a BLAGN similar to PDS~456 with $\Gamma=1.8$ and $k_{\rm bol,X}=10$ (green) and (ii) an NLS1 AGN similar to Ark~564 with a soft excess, $\Gamma=2.4$ and $k_{\rm bol,X}=30$ or $100$ (dashed red). The shaded region indicates the range of X-ray emission for models with a Balmer-break strength $D_{4000}=2.5$ and a metallicity $Z/Z_\odot\in[0.05;0.1]$. A Galactic column density of $N_\mathrm{H}=10^{21} \,\mathrm{cm^{-2}}$ is assumed for the intrinsic X-ray templates, while the Balmer-break strength $D_{4000}=2.5$ corresponds to a column density of $N_\mathrm{H}\!\sim\!10^{25} \,\mathrm{cm^{-2}}$. Both intrinsic X-ray templates exceed the observational upper limits. After attenuation by the cocoon, however, the BLAGN X-rays are still too bright to be consistent with the data, whereas the softer NLS1 template is easily consistent with the observed upper limits.
}
\label{fig:compton_thick}
\end{figure}

\section{Model}\label{sec:model}

The X-ray detectability of LRDs is set by the intrinsic coronal spectral energy distribution (SED) and by the transmission through the surrounding gas cocoon. We parametrised the intrinsic emission in terms of the bolometric luminosity ($L_{\rm bol}$), the X-ray bolometric correction (\kbolx), and the shape of the X-ray SED, while the attenuation was described by the X-ray transmission factor, $T_\mathrm{X}$. The bolometric luminosities of LRDs are increasingly constrained by large samples that map the diversity of the population \citep{deGraaff2025b,Barro2025,PerezGonzalez2026} and by high-fidelity radiative transfer that reproduces their observed spectra. We computed $T_\mathrm{X}$ from the gas-cocoon properties inferred from such modelling. 
We adopted the LRD model sequences of \citet{Sneppen2026}, computed with the Monte Carlo radiative-transfer code \sirocco \citep{long2002,Matthews2025}. In these models, an increasing gas column density produces stronger Balmer breaks, here expressed as $D_{4000}=F_{\lambda=4200}/F_{\lambda=3500}$, and sets the width of the electron-scattering line wings. As we quantify below, larger columns also progressively attenuate the X-ray emission. The mapping between $D_{4000}$ and column density is non-linear, because when the gas becomes optically thick short-ward of the Balmer break, the transmitted flux there declines rapidly with increasing column. Throughout this analysis, we assumed that $D_{4000}$ traces the column density, motivated by the modest dynamic range of observed $D_{4000}$ and the presence of UV emission lines from LRDs, such as broad Ly$\alpha$ \citep{Ji2026,Tang2025}. If part of the UV emission instead arises outside the LRD itself, the cocoon could have a stronger Balmer break and a higher column density, allowing a higher intrinsic X-ray luminosity. For the X-ray calculations, we used the fiducial parameters, radiative-transfer model, and cocoon ionisation structure from the same model grid. The main change was that we ran \sirocco in classic mode with weight reduction, rather than in indivisible-packet macro-atom mode; in the current implementation, this allowed us to treat Compton heating and cooling more accurately.

We anchored the adopted $L_{\rm bol}$ to the observed rest-frame optical emission, where much of the reprocessed luminosity emerges \citep[e.g.][]{Greene2025}, and not to an unobscured-AGN bolometric correction. A higher $L_{\rm bol}$ would rescale the inferred lower limit on $k_{\rm bol,X}$ linearly, but leave unchanged the transmission $T_X$ and its dependence on column density, metallicity, redshift, and intrinsic X-ray SED. Fig.~\ref{fig:compton_thick} shows the intrinsic X-ray template spectra for a broad-line AGN (BLAGN) and a narrow-line Seyfert 1 (NLS1) on the assumption that they are little altered in such objects after leaving the innermost \(10^{16}\)\,cm or so. We adopted smooth continuum shapes as intrinsic templates, with fiducial parameters motivated by AGN population properties and well-studied objects with high accretion rates, PDS~456 and Ark~564 
\citep{Kara2017,Reeves2021}. We normalised the templates using the X-ray bolometric correction $k_{\rm bol,X}=L_{\rm bol}/(\nu L_{2-10\,\mathrm{keV}})$ and explored variations in \kbolx. \kbolx$\approx10$ is consistent with bolometric corrections for low-Eddington AGNs \citep[e.g.][]{Vasudevan2007}, while \kbolx can increase to $\sim\!100$ at near Eddington accretion rates \citep{Lusso2012} and may be even higher if the accretion disk quenches the corona \citep{Proga2005}. 
The expected hard power-law photon indices, $N(E)\propto E^{-\Gamma}$, for NLS1s and BLAGNs are $\Gamma_{\rm NLS1}=2.1\pm0.3$ \citep[e.g.][]{Barua2020,Yu2023} and $\Gamma_{\rm BLAGN}\approx1.8$ \citep[e.g.][]{Ricci2018}. The high-energy cut-off of this power law is $E_{\rm C,NLS1}\sim30{-}60\,$keV \citep{Malizia2008} and $E_{\rm C,BLAGN}=210\pm36\,$keV \citep{Ricci2018}. In the Compton-thick regime relevant for the LRD cocoon, harder intrinsic spectra provide more high-energy photons that can be down-scattered into the observed bandpass. They therefore predict a higher observed flux after reprocessing. The least stringent limits on the intrinsic X-ray emission are thus obtained for the softest assumed input spectrum, corresponding to the steepest power-law index and the lowest cut-off energy. Our least constraining input X-ray SED was therefore $\Gamma_{\rm NLS1}=2.4$ and $E_{\rm C,NLS1}=30\,$keV. 

The X-ray templates were used to bracket the intrinsic coronal SED before propagation through the cocoon. For each assumed intrinsic spectral shape and cocoon composition, the X-ray upper limits impose a lower limit on \kbolx. We then compared this inferred \kbolx with empirical AGN population relations, in which higher \kbolx values are typically associated with higher Eddington ratios \citep[e.g.][]{Duras2020,Gupta2025}. Thus, the inference is that normal hard X-ray-bright coronal spectra are difficult to hide, while soft X-ray-weak spectra characteristic of high-accretion AGNs remain viable.

\begin{figure*}
\vspace{-0.1cm}
\begin{center}
    \includegraphics[angle=0,width=\textwidth]{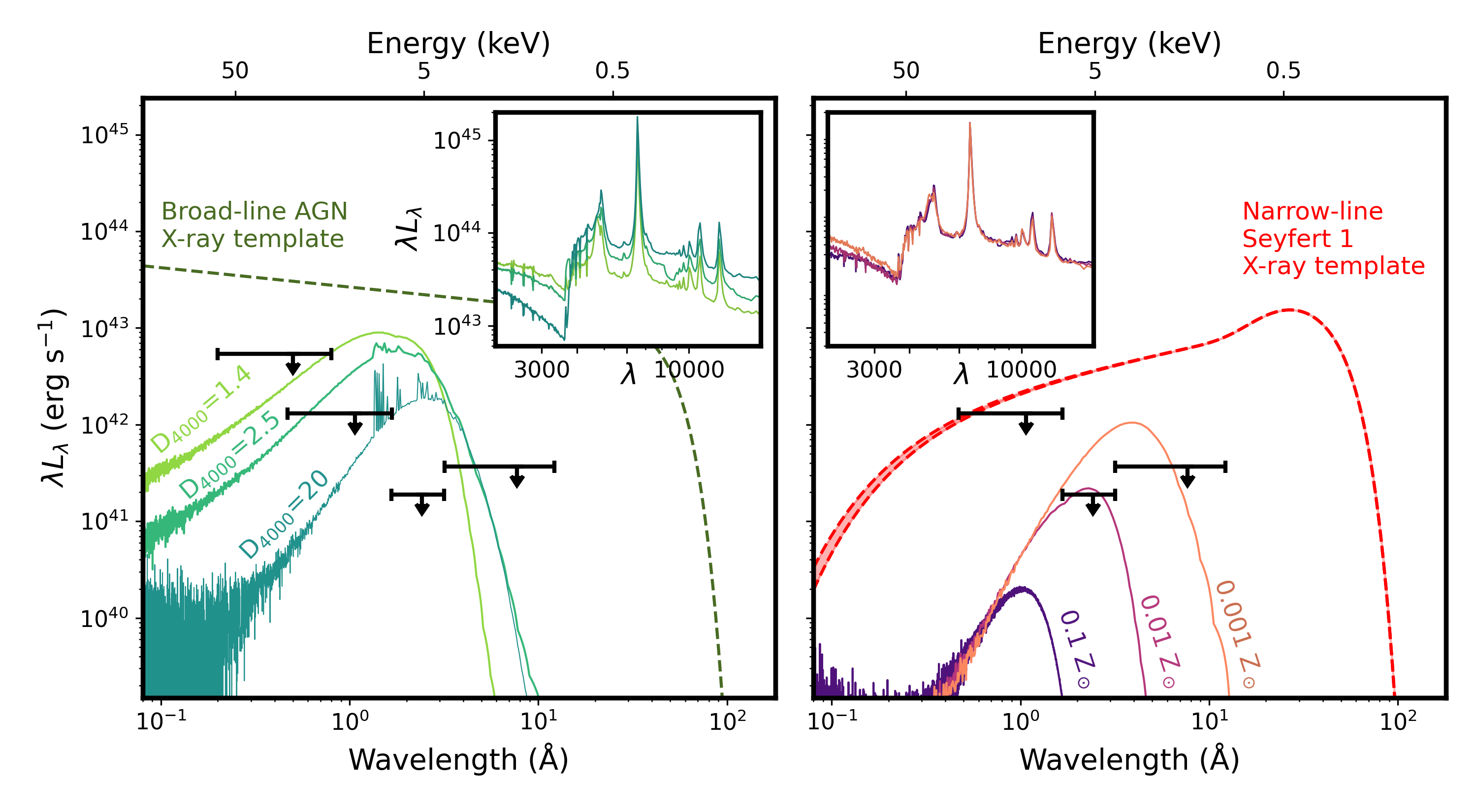}
\end{center}
\vspace{-0.8cm}
\caption{X-ray Compton peaks and UV--near-IR spectra (\emph{inset}) for LRD gas cocoons with varying column density (left) and gas composition (right). In the left panel, $0.01Z_\odot$, $k_{bol,X}=10$ and a BLAGN X-ray template is assumed; in the right panel, $D_{4000}=2.5$, $k_{bol,X}=50$, and an NLS1 X-ray template are assumed. Changes in composition only modestly affect the rest-frame optical spectrum via the plasma conditions, but they strongly regulate the photoelectric absorption of soft X-rays. Very low-metallicity cocoons, as expected for pristine direct-collapse BH formation conditions, provide much lower X-ray attenuation.}
\label{fig:metallicity}
\end{figure*}

\section{Results}\label{sec:results}
For the Balmer-break strengths typical of most LRDs, $D_{4000}\in1{-}10$ \citep{deGraaff2025b,Sneppen2026}, the implied column densities are $N_\mathrm{H}\!\sim\!10^{24.5}{-}10^{25.5} \,{\rm cm^{-2}}$. At these column densities, the emergent X-ray spectra are characterised by a Compton hump reminiscent of Compton-thick AGNs \citep{George1991,Magdziarz1995,Comastri2004} and simple X-ray models \citep{Murphy2009,Brightman2011}: multiple Compton scatterings redistribute higher-energy photons to lower energies, while photoelectric absorption suppresses the soft X-rays (see Fig.~\ref{fig:compton_thick}). 
Compton scattering traces the free-electron column and is dominated by the highly ionised gas interior to the Balmer-break-forming layer. Photoelectric absorption, by contrast, is dominated by bound inner-shell electrons in metals and is therefore important anywhere the gas is not fully stripped, that is, along most of our column, including but not limited to the Balmer-continuum opacity region.
With the magnitude of the Balmer breaks actually observed, the BLAGN template predicts X-ray fluxes that are above the observational limits, whereas the softer and less luminous NLS1 template lies below (i.e.\ consistent with) them. 

In Fig.~\ref{fig:metallicity} we illustrate the parameters that shape the X-ray emission, namely the intrinsic X-ray SED, the column density, and the gas composition.
An increasing column density increases the Balmer-break strength and shifts the X-ray emission to lower energies through enhanced Compton scattering. However, the relation between the X-ray suppression and UV attenuation as inferred from the Balmer break is complex and non-monotonic. Larger Balmer breaks do not necessarily imply weaker X-ray flux at a fixed energy below the Compton hump as Compton down-scattering redistributes photons into lower-energy bands. 
As a result, depending on the bandpass considered and the metallicity of the gas, the X-ray suppression, $T_\mathrm{X}$, can be either greater or lower than the UV attenuation suggested from the Balmer-break strength. 

The cocoon composition strongly affects the degree of photoelectric absorption. An order-of-magnitude increase in metallicity leads to an increase of a factor of $\sim2$ in the Compton peak energy and to a decrease of a factor of $\sim3$ in the peak transmitted fraction. These emergent spectral properties depend non-linearly on metallicity because the transmitted spectrum is shaped by exponential photoelectric attenuation and Compton reprocessing. The underlying X-ray photoelectric optical depth, $\tau_\mathrm{X}$, however, scales approximately linearly with metallicity because absorption at the relevant X-ray energies is dominated by metals. Equivalently, the solar-abundance-equivalent X-ray column density is $N_{\mathrm{H_X}(Z)} = (\tau_\mathrm{X}/\tau_\mathrm{X_\odot}) N_\mathrm{H} \sim (Z/Z_\odot) N_\mathrm{H}$. 
Very low metallicities, either pristine formation values $Z/Z_\odot\lesssim10^{-5}$ invoked in direct-collapse BH scenarios \citep[e.g.][]{Omukai2008,Cenci2025}, or enriched values $Z/Z_\odot\lesssim5\times10^{-3}$ \citep{Pacucci2026}, imply weak photoelectric absorption and therefore only modest attenuation in the observed-frame 0.5--5\,keV band. The lack of detection in X-rays therefore suggests that the gas in LRDs is not extremely metal poor when LRDs have an NLS1 or a brighter AGN at their cores. 

No spectroscopically confirmed LRD, defined by the presence of a Balmer break and broad lines, has yet been significantly detected in X-rays, although a few phenomenologically similar compact red sources have been detected in X-rays \citep{Fu2025,Hviding2026,Kocevski2024}. The deep-field limits from \emph{Chandra} are stringent, with a combined limit of $L_{\rm 1{-}10\,keV}\lesssim(2{-}4)\times10^{41}\,{\rm erg\,s^{-1}}$ \citep{Maiolino_Chandra_2024,Comastri2026} on classical LRDs. The Rosetta Stone LRD alone has a rest-frame X-ray limit of $L_{\rm 2{-}10\,keV}<2.9\times10^{41}\,{\rm erg\,s^{-1}}$, to be compared with a bolometric luminosity higher by three orders of magnitude at $\sim3\times 10^{44}\,{\rm erg\,s^{-1}}$, despite a modest Balmer break, $D_{4000}\sim2$. This combination is examined in Fig.~\ref{fig:compton_thick}, where a gas cocoon with a column density of $N_H=10^{25}\,{\rm cm^{-2}}$ produces a Balmer break with a strength similar to that of the Rosetta Stone object in the \sirocco models. In this model, a BLAGN X-ray template with $k_{\rm bol,X}=10$ transmits enough X-rays to contravene the observational upper limits. By contrast, a narrow-line Seyfert~1 X-ray template is consistent with the upper limits for standard bolometric corrections $k_{\rm bol,X}\sim10{-}100$. This conclusion is conditional on the assumed intrinsic X-ray SED and cocoon properties, but within our fiducial models, the key dependence is the metallicity of the absorbing gas. The limited LRD host metallicities from narrow-line ratios suggest metallicities in the range $\sim\!0.02$-$0.2\,Z_{\odot}$ \citep{Juodzbalis2024_rosetta,Ivey2026,Nikopoulos2026}. For an NLS1-like X-ray template and metallicities representative of the lower-metallicity part of the population, $\sim0.04Z_{\odot}$, the limits imply $k_{\rm bol,X}\gtrsim30$. Such high bolometric corrections are within the scatter of standard AGN relations, but at the population-level would suggest Eddington ratios near unity; $\gtrsim 0.1 \,L_{\rm edd}$ \citep{Vasudevan2007}, $\gtrsim0.3 \,L_{\rm edd}$ \citep{Gupta2025} and/or $\gtrsim0.6 \,L_{\rm edd}$ \citep{Duras2020}. A metallicity as low as $\lesssim0.01\,Z_\odot$ would shift the Compton peak into the tightly constrained soft X-ray band (Fig.~\ref{fig:metallicity}), requiring $k_{\rm bol,X}\gtrsim100$, which is seen rarely \citep{Martocchia2017}. 
Local LRD analogues \citep{Lin2024,Lin2026} are not necessarily more constraining at moderate metallicity because their observed soft X-ray band probes strongly photoelectrically absorbed rest-frame energies; they would, however, strongly test extremely metal-poor cocoons (Fig.~\ref{fig:distance}).

\section{Predictions and conclusions}\label{sec:predictions}

We have tested whether intrinsic AGN X-ray templates are compatible with the current X-ray upper limits for LRDs. The gas columns required to reproduce their optical and near-infrared SEDs naturally produce substantial X-ray attenuation, but in our fiducial models, this attenuation is generally insufficient to hide a normal hard X-ray-bright corona. Metallicity remains a major uncertainty in the degree of soft X-ray flux suppression, while the hard X-ray emission at rest-frame energies $\gtrsim10\,{\rm keV}$ is governed mainly by the column density and the intrinsic X-ray SED. Non-detections in the \emph{Chandra} deep fields even in circumstances with favourable redshift and bandpass, low metallicities, and modest columns, suggest that (i) the intrinsic X-ray SEDs of LRDs are soft and/or (ii) LRDs are intrinsically weak X-ray sources relative to their bolometric luminosities \citep[e.g.][]{Duras2020}, consistent with high-accretion-rate AGNs. Compared to the higher-energy Compton hump of standard AGNs, which peak at $\sim\!10{-}30$\,keV \citep{Zhu2021}, the LRD model X-rays can peak at lower energies of 4--7\,keV due to the combination of high densities, which enhance Compton scattering, and lower metallicities, which reduce photoelectric absorption. The \sirocco modelling framework makes several predictions for X-ray detectability. 
\begin{enumerate}
    \item Mild Balmer-break LRD systems, particularly at low metallicities with $\lesssim 0.001{-}0.05Z_{\odot}$, exhibit only mild soft X-ray suppression, as in Little Blue Dots \citep{Sneppen2026d}, and they are therefore promising targets for X-ray searches. Systems with small $D_{4000}$ and low metallicities (e.g.\ Jades-GN 1181-73488; \cite{Nikopoulos2026}) are particularly attractive candidates. We note that this object has a marginal X-ray detection \citep{Maiolino_Chandra_2024}.
    \item Objects with extremely low metallicities and moderate column densities, such as those invoked for a direct-collapse BH interpretation of LRDs with $Z\ll10^{-2}Z_{\odot}$ and a few $10^{25} \rm cm^{-2}$ \citep{Pacucci2026}, are not strongly suppressed in the \emph{Chandra} bands. For typical \kbolx, the X-ray non-detections therefore disfavour pristine absorbing gas. 
    \item A population of LRDs with very strong Balmer breaks and narrow electron-scattering wings are expected to be extremely X-ray weak: the high columns required to reprocess the broad wings, $N_\mathrm{H}\gtrsim5\times10^{25}\,{\rm cm^{-2}}$, can suppress X-rays even at low metallicity.
    \item Although LRDs can be labelled type~1 AGNs due to their broad lines, the broad wings may instead be produced by electron scattering and not by an unobscured broad-line region. In this case, the broad lines are evidence of high scattering columns, which is consistent with strong obscuration in the UV and X-rays and not evidence against it.
    \item Standard AGN luminosity relations $L_{\rm2\,keV}{-}L_{\rm2500\,Å}$ \citep{Lusso2010} are thrown off in LRDs because dense gas can attenuate the UV and X-ray emission both, which easily introduces a systematic shift of more than an order of magnitude and scatter in the relations between these luminosities. 
    For the same reason, bolometric corrections calibrated on unobscured local AGNs are unlikely to be reliable. Comparisons to empirically inferred or radiative-transfer-based bolometric luminosities or to the optical luminosity, where much of the reprocessed emission appears to emerge \citep{Greene2025}, are likely to be more informative.
\end{enumerate}
We emphasise that X-ray non-detections alone cannot rule out a corona-free accretion state (which would be the ultimate \kbolx). Our constraints therefore apply primarily to AGN-like coronal templates: for such input spectra, the cocoon columns inferred from the optical spectra predict some X-ray weakness due to attenuation, but this is generally insufficient to hide the X-ray output of normal BLAGNs.
%In the absence of strongly

\begin{acknowledgements}
The authors would like to thank Giorgos Nikopoulos for helpful discussions. AS, DW, \& SAS are funded in part by the European Union (ERC, HEAVYMETAL, 101071865). Views and opinions expressed are, however, those of the authors only and do not necessarily reflect those of the European Union or the European Research Council. Neither the European Union nor the granting authority can be held responsible for them. 
JHM acknowledges funding from a Royal Society University Research Fellowship (URFR1221062). SAS is supported by the UK’s Science and Technology Facilities Council (STFC, respectively grant ST/V001000/1 and ST/X00094X/1). 
\end{acknowledgements}

\bibliographystyle{aa}
\bibliography{refs} 
\begin{appendix}

\renewcommand{\thefigure}{A.\arabic{figure}}

\begin{figure}
\begin{center}
    \includegraphics[angle=0,width=0.5\textwidth]{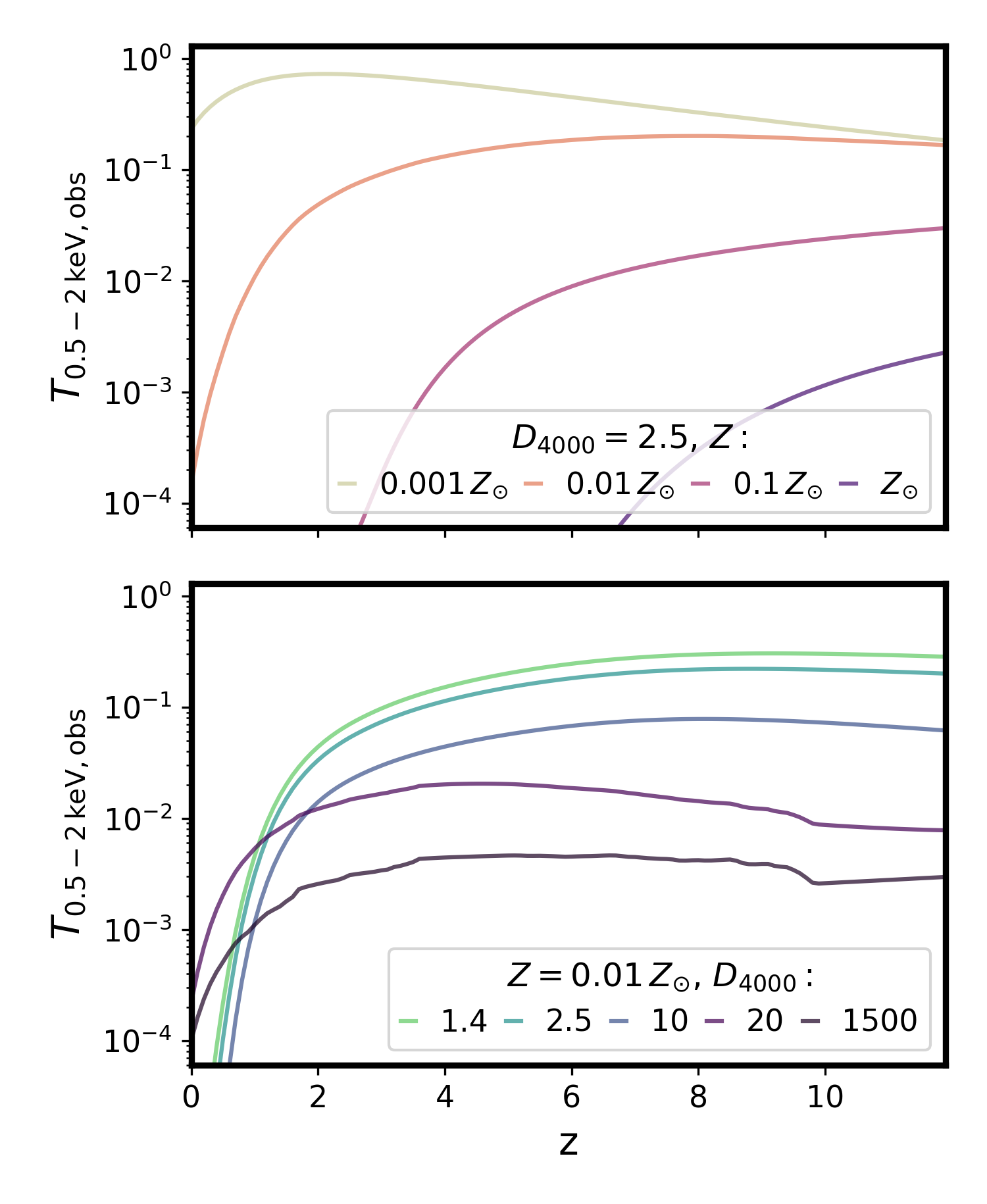}
\end{center}
\vspace{-0.6cm}
\caption{X-ray transmitted fraction, $T_{\rm 0.5-5\,keV,obs}$, as a function of redshift, metallicity, and cocoon column density, parametrised by the Balmer-break strength $D_{4000}$. The highest transmission occurs for low metallicity, modest redshift, and weak Balmer breaks, whereas metal-enriched cocoons at low redshift strongly suppress X-rays by photoelectric absorption. In the most optically thick regimes, high ionisation X-ray emission lines begin to dominate the transfer function, which produces `wiggles' as the bandpass moves over ionised lines.
}
\label{fig:trans_fraction}
\end{figure}

\begin{figure}
\begin{center}
    \includegraphics[angle=0,width=0.5\textwidth]{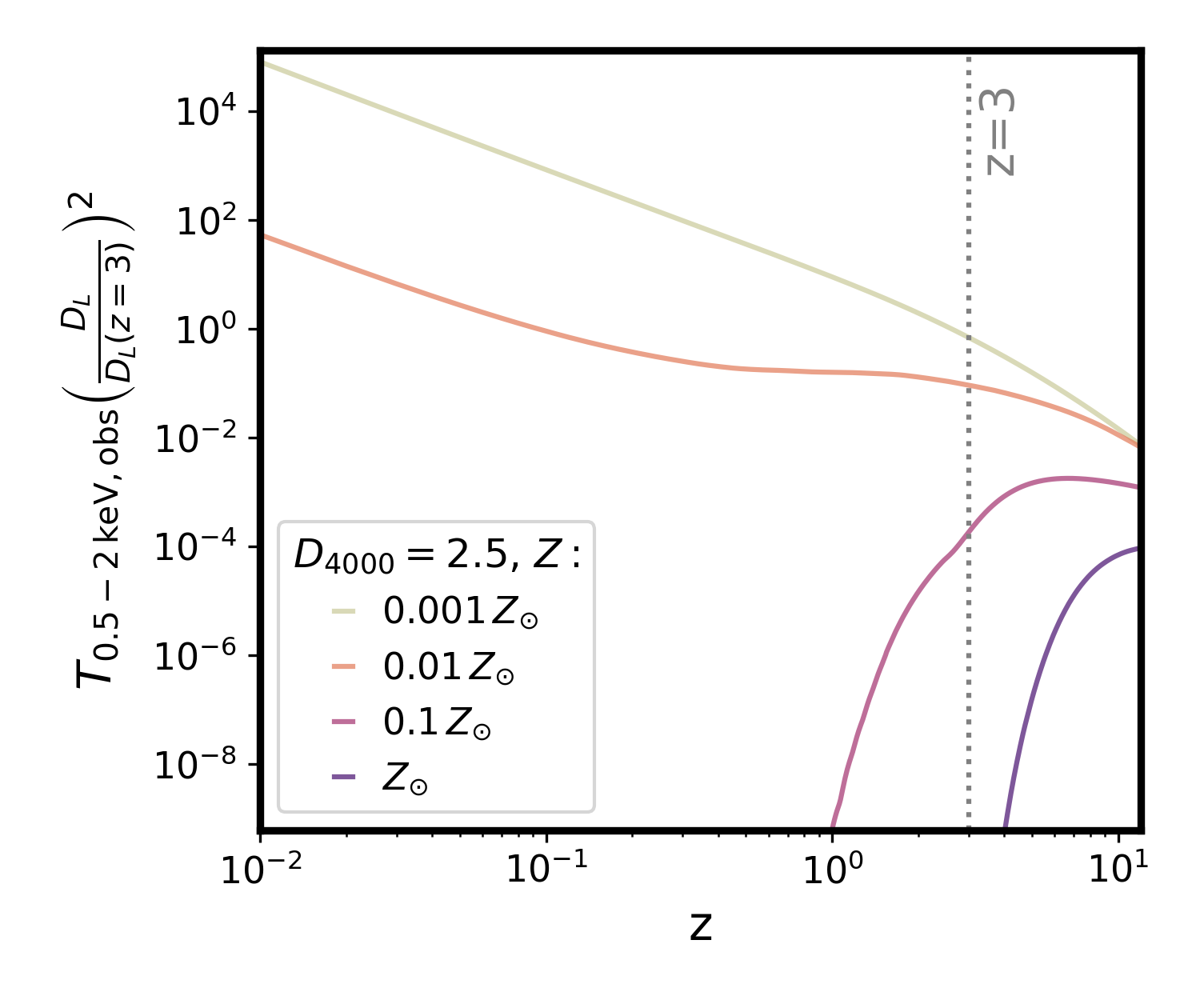}
\end{center}
\vspace{-0.6cm}
\caption{X-ray transmitted fraction, $T_{\rm 0.5-5\,keV,obs}$, weighted by flux dilution with increasing distance (e.g. by luminosity distance $D_L^2$). Notably, local LRD analogues with low-redshift ($z\lesssim2$) and moderate metallicity ($Z\sim0.1Z_\odot$) systems are in the highly suppressed photo-electric absorption band.  }
\label{fig:distance}
\end{figure}

\begin{figure}
\begin{center}
    \includegraphics[angle=0,width=0.5\textwidth]{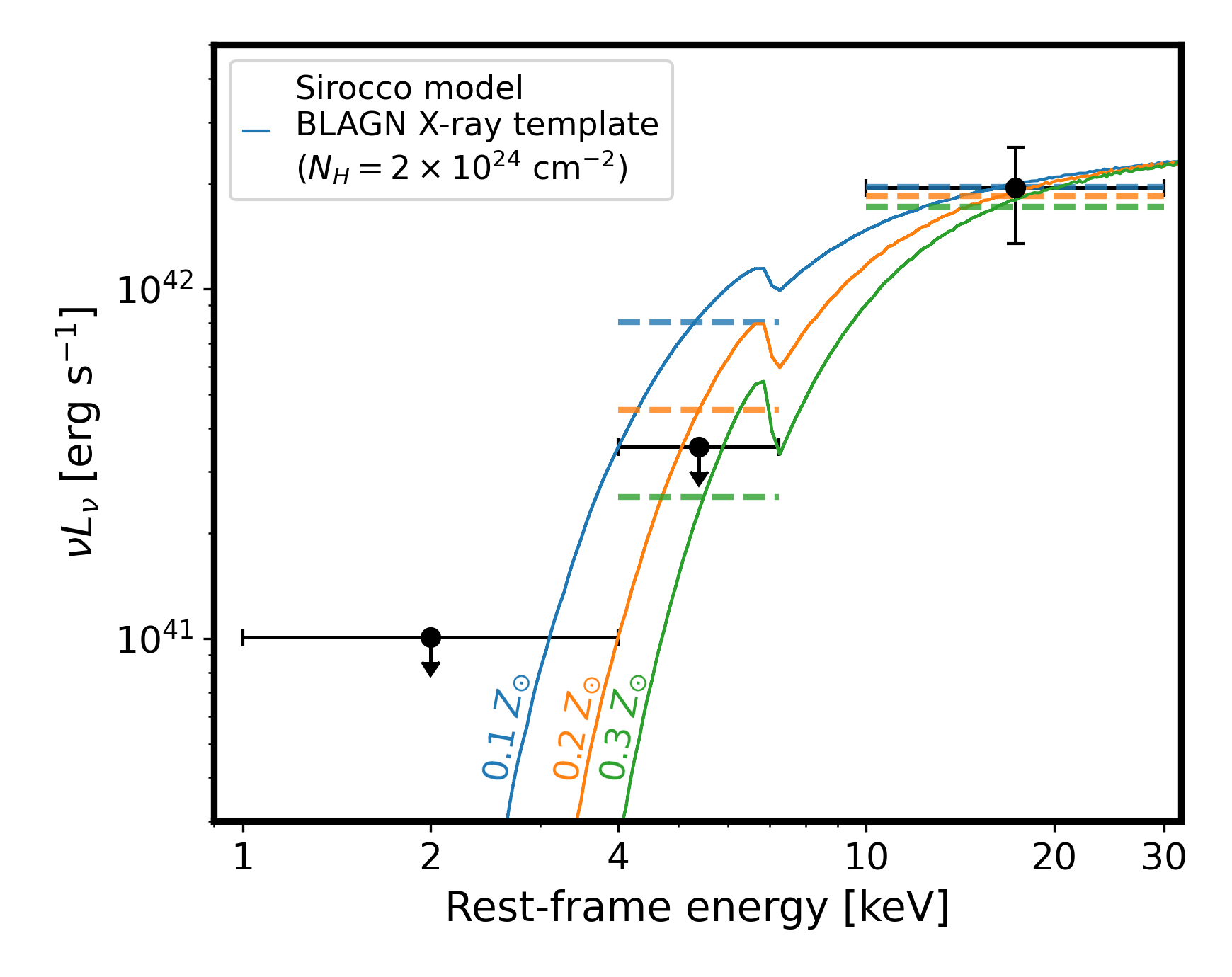}
\end{center}
\vspace{-0.6cm}
\caption{Type 2 JWST AGN X-ray upper limits and high-energy band detection from \cite{Comastri2026} alongside \sirocco emergent X-ray spectra. Similar to the inferred modest Compton-thick columns of $2\times10^{24}\,{\rm cm^{-2}}$ in \cite{Comastri2026}, we can reproduce the observed hard X-ray spectrum for sufficient photoelectric absorption ($Z\gtrsim0.2Z_{\odot}$).  }
\label{fig:type2}
\end{figure}

\section{Atomic data}\label{app:atomic}

The X-ray photoelectric opacity is computed using the \sirocco atomic data set described by \citet{Matthews2025}, which contain photoionisation cross-sections, including those for inner-shell cross-sections relevant for X-ray absorption \citep{Cunto1993,Verner1993,Verner1995,Verner1996}. Additional details on X-ray signatures in Monte Carlo radiative transfer and on the applicability of \sirocco to X-ray radiative transfer can be found in \citet{Sim2008} and \citet{Higginbottom2013}, respectively.

\section{Band-integrated transmission fraction}\label{app:transmission}

In Fig.~\ref{fig:trans_fraction} we quantify the band-integrated transmission fraction as a function of the redshift, $z$, for different metallicities and cocoon column densities. The conditions for minimal X-ray suppression occur at intermediate redshifts, $z\sim2{-}3$, low metallicities, $Z\lesssim0.1$, and low column densities ($D_{4000}\sim1$; e.g. SEDs with continuum inflection near the Balmer break wavelength). In fact, Compton down-scattering can even redistribute photons into the soft bands and increase the soft-band luminosity relative to the intrinsic emission.
Lower-redshift objects can be strongly obscured by photoelectric absorption because we are observing their softer X-rays. Fortunately, objects within the \emph{Chandra} Deep Fields, such as Jades-GN 1181-73488 \citep{Juodzbalis2025} and the Rosetta Stone LRD \citep{Juodzbalis2024_rosetta}, broadly fulfil such criteria and are relatively bright, providing useful case studies for comparison with deep observational limits. 

In this context, it is worth emphasising that the softer bandpass probed by local LRD analogues suggests strong expected photoelectric absorption (see Fig.~\ref{fig:distance}). In particular, the combination of low redshifts and modest metallicities $\sim0.13\,Z_{\odot}$ \citep{Lin2024,Lin2026} implies nearby sources at $z\sim0.1$-$0.2$ may be X-ray fainter than the population at $z\sim3$ in the high-sensitivity observed soft X-ray bands. Conversely, assuming they are chemically pristine (e.g. $\lesssim10^{-2}-10^{-3}\,Z_{\odot}$) it would be very difficult to hide X-ray emission from such local LRDs. The strong sensitivity to redshift and bandpass arises because the photoelectric cross-section $\sigma_{\rm pe}\propto E^{-3}\propto\lambda^3$ away from absorption edges, making the transmitted flux fall roughly as $\propto e^{-\lambda^3}$.

\section{Bolometric corrections}\label{app:bol}

The bolometric luminosity of LRDs remains uncertain, but empirical constraints suggest that this uncertainty is not well described by standard unobscured-AGN bolometric corrections. In particular, X-ray-to-FIR SED integrations of luminous LRDs indicate that more than half of the emergent luminosity may be emitted in the rest-frame optical \citep{Greene2025}. This is consistent with dense-gas models as photons shortward of the optical are efficiently reprocessed, while current MIRI/ALMA constraints disfavour a large, ubiquitous hidden dust-reprocessed component \citep{Setton2025,Casey2025}. Similarly, radiative-transfer modelling can reproduce the optical spectra without requiring reddening needed for a bolometrically dominant dust component \citep{Sneppen2026}. In the main text, we have therefore anchored $L_{\rm bol}$ to the observed rest-frame optical emission, which is conservative with respect to any hidden luminosity component. For fixed cocoon transmission and X-ray upper limit, any increase in $L_{\rm bol}$ rescales the inferred lower limit on $k_{\rm bol,X}$ linearly. Thus, the conclusion that Compton-thick gas alone does not generally hide a normal hard, X-ray-bright corona is robust to, and would be strengthened by, larger $L_{\rm bol}$, since this would require even more extreme $k_{\rm bol,X}$.

\section{Type 2 X-ray detection}\label{app:type2}

Beyond the LRD X-ray upper limits, \cite{Comastri2026} found a $\sim\!3\sigma$ detection for the type 2 JWST AGN population in the hardest rest-frame band ($10$-$30$ keV). Given non-detection in softer rest-frame bands ($1$-$4$ keV and $4-7.25$ keV), they interpreted the stacked spectrum using solar-abundance \textsc{uxclumpy} obscuration templates, finding it consistent with Compton-thick columns, $N_H\gtrsim2\times10^{24}\,{\rm cm^{-2}}$. We recover a similar inference for this X-ray detection as illustrated by the curves in Fig.~\ref{fig:type2} under the assumption of sufficient photoelectric absorption (i.e. $\gtrsim0.2Z_{\odot}$). Here the model spectra use as input the mean bolometric luminosity of the type 2 AGN sample \citep[$6\times10^{43} \,{\rm erg/s}$,][]{Scholtz2025}, $k_{\rm bol,X}=10$ and the example BLAGN X-ray template. For comparison with spectral models, we plot each reported band luminosity as a single representative $\nu L_\nu$ error bar by dividing by $\ln(E_2/E_1)$, such that a flat $\nu L_\nu$ curve across the band would give the reported $L_{\rm band}$. 
We can also match the detection with the narrow-line Seyfert~1 X-ray template and a higher transmitted X-ray fraction, but importantly the key constraint provided by the hard/soft X-ray ratio is on the photoelectric absorption (e.g. metallicity of the gas), which must be substantial given the dearth of soft X-rays. Notably, a value of $\gtrsim0.2Z_{\odot}$ hints at higher metallicities in the probed type 2 population, than is typically inferred for the LRD population \citep{Juodzbalis2024_rosetta,Ivey2026,Nikopoulos2026}. 
\end{appendix}

\end{document}